\documentclass[sigconf]{acmart}

\usepackage{booktabs} % To thicken table lines

\usepackage[english]{babel}
\usepackage{moresize}
\usepackage{amsmath}
\usepackage{algorithmic} 
\usepackage{balance}
\usepackage{comment}
\usepackage{paralist}
\usepackage{bm}
\usepackage{pgfplots}
\usetikzlibrary{pgfplots.dateplot}

\usepackage{flushend}
\usepackage[english]{babel}
\usepackage[latin1]{inputenc}
\usepackage{mathrsfs}
\usepackage{graphicx}

\usepackage{amsfonts}
\usepackage{url}
\usepackage{longtable}
\usepackage{rotating}
\usepackage{multirow}
\usepackage{mathrsfs}
\usepackage{subfigure}
\usepackage{enumitem}
\usepackage[linesnumbered,algoruled,boxed,lined]{algorithm2e}
\usepackage{adjustbox}
\usepackage{hyperref}
\usepackage{balance}
\usepackage{pgfplots}
\usetikzlibrary{pgfplots.dateplot}

\usepackage{filecontents}
\usepackage{lineno} % number the lines

\definecolor{tblue}{RGB}{31,119,180}
\definecolor{torange}{RGB}{255,127,14}
\definecolor{tgreen}{RGB}{44,160,44}
\definecolor{tred}{RGB}{214,39,40}
\definecolor{tpurple}{RGB}{148,103,189}

\newcommand{\hide}[1]{} %hide

\newcommand{\etal}{\textit{et al}.}

\newcommand{\ie}{\textit{i}.\textit{e}.}
\newcommand{\eg}{\textit{e}.\textit{g}.}

\setcopyright{acmlicensed}
\def\model{SMIN}

\copyrightyear{2021}
\acmYear{2021}
\setcopyright{acmlicensed}\acmConference[CIKM '21]{Proceedings of the 30th ACM International Conference on Information and Knowledge Management}{November 1--5, 2021}{Virtual Event, QLD, Australia}
\acmBooktitle{Proceedings of the 30th ACM International Conference on Information and Knowledge Management (CIKM '21), November 1--5, 2021, Virtual Event, QLD, Australia}
\acmPrice{15.00}
\acmDOI{10.1145/3459637.3482480}
\acmISBN{978-1-4503-8446-9/21/11}

\begin{document}

\fancyhead{}

\author{Xiaoling Long$^{1*}$, Chao Huang$^{2*}$, Yong Xu$^{1}$, Huance Xu$^2$, Peng Dai$^3$, \\ Lianghao Xia$^1$, Liefeng Bo$^3$}
\thanks{The first two authors contribute equally to this work}
\thanks{Corresponding author: Yong Xu}
\affiliation{$^1$South China University of Technology, $^2$University of Hong Kong, $^3$JD Finance America Corporation\\
\{mslongxiaoling, cshuance.xu, cslianghao.xia\}@mail.scut.edu.cn, \\chaohuang75@gmail.com, yxu@scut.edu.cn, \{peng.dai, liefeng.bo\}@jd.com
}

% \title{Global Context Enhanced Social Recommendation via Graph Mutual Information Maximization}
\title{Social Recommendation with Self-Supervised Metagraph Informax Network}

\begin{abstract}
In recent years, researchers attempt to utilize online social information to alleviate data sparsity for collaborative filtering, based on the rationale that social networks offers the insights to understand the behavioral patterns. However, due to the overlook of inter-dependent knowledge across items (\eg, categories of products), existing social recommender systems are insufficient to distill the heterogeneous collaborative signals from both user and item side. In this work, we propose Self-Supervised Metagraph Informax Network (\model) which investigates the potential of jointly incorporating social- and knowledge-aware relational structures into the user preference representation for recommendation. To model relation heterogeneity, we design a metapath-guided heterogeneous graph neural network to aggregate feature embeddings from different types of meta-relations across users and items, empowering \model\ to maintain dedicated representations for multifaceted user- and item-wise dependencies. Additionally, to inject high-order collaborative signals, we generalize the mutual information learning paradigm under the self-supervised graph-based collaborative filtering. This endows the expressive modeling of user-item interactive patterns, by exploring global-level collaborative relations and underlying isomorphic transformation property of graph topology. Experimental results on several real-world datasets demonstrate the effectiveness of our \model\ model over various state-of-the-art recommendation methods. We release our source code at https://github.com/SocialRecsys/SMIN.

% Further analysis provides insights into the performance superiority of our new recommendation framework. 

%and show the benefits of user/item representations brought by \model's incorporation of 

% recommendation performance since item knowledge graph signals could serve as important hetergourous 
\end{abstract}



\maketitle

\section{Introduction}
\label{sec:intro}

Recommender systems have become an essential component in online platforms (\eg, E-commerce~\cite{huang2019online}, news portals~\cite{wu2019npa}) to alleviate the information overload for users~\cite{2020learning,wang2019neural}. To predict user preference over different items from historical behaviors, collaborative filtering (CF) has served as the key paradigm to make recommendation based on users' potential common preferences~\cite{wang2019unified,chen2020efficient}. However, conventional CF methods often suffer from data sparsity and cold start problems. With the prevalence of online social communities in enabling users to share their opinions with others, many efforts have been devoted to incorporating online social network information into the user-item interaction learning, to alleviate the data sparsity issue and enhance recommendation performance~\cite{fan2019graph,wu2019neural,chen2019social}.

These social recommendation models are proposed based on the social influence theory that socially connected people are likely to share similar interests~\cite{lewis2012social}. Earlier methods extend the matrix factorization architecture to integrate social relationships as regularization term (\eg, SoReg~\cite{ma2011recommender}) or project users into latent representations with their trust relationships (\eg, TrustMF~\cite{yang2016social}). In recent years, several efforts provide deep insights into neural network techniques and design more sophisticated models to learn precise user and item embeddings, with the joint integration of user-user and user-item relational structures. In particular, attention-based mechanisms, such as SAMN~\cite{chen2019social} and EATNN~\cite{chen2019efficient}, perform the attentive operations over relationships between users through weighted embedding summation. Furthermore, due to the graph-structured nature of users' social connections, several recent works design graph neural social recommendation models to encode the user relation graph into latent representation space. For instance, GraphRec~\cite{fan2019graph} and DANSER~\cite{wu2019dual} adopt the graph attention network to aggregate feature embeddings from neighboring nodes of the target user. Motivated by the strength of graph convolution network, DiffNet~\cite{wu2019neural} is built upon the graph-structured message passing architecture to conduct the layer-wise information diffusion across users and approximate the social influence propagation process.

Despite the aforementioned solutions have provided promising recommendation performance, we argue that existing social-aware recommender systems are insufficient to yield quality representations for preserving user's preference, since their designed embedding functions lack an explicit encoding of semantic relatedness between items~\cite{yu2018walkranker}. In real-world recommendation applications, incorporating the external knowledge from item side is particularly helpful to better understand user's interests~\cite{wang2019multi,xin2019relational}. For example, in online retail platforms, products can be correlated with their similar functionality~\cite{xu2020product}. External knowledge from online review systems (\eg, Yelp) may contain fruitful connections between venues in terms of their categorical similarities or received reviews~\cite{zhang2020aser}. These semantic signals have great potential to characterize knowledge-aware relationships across different items~\cite{zhou2020interactive}, which can enhance the collaborative effect encoding for social recommendation. 

The joint modeling of social- and knowledge-aware collaborative signals poses several unique technical challenges that cannot be easily handled by existing social recommendation methods. \textbf{First}, the incorporation of both social and knowledge graph information into user-item interactions, involves diversified inter-dependencies which are heterogeneous in nature. How to effectively preserve the heterogeneous relational structure across users and items in a unified representation framework, remains a significant challenge. \textbf{Second}, in practical recommendation scenarios, factors that affect customer's behavior are often multifaceted from both user and item domains~\cite{wu2019dual,wang2020modeling}. For example, there exist multiple relations between items reflecting multi-dimensional contextual signals, \eg, behavior-level co-interact patterns and knowledge graph-based item-wise dependencies. Hence, it is a necessity to endow the designed embedding function with the capability of capturing item-item semantic relatedness and knowledge-aware social influence. \textbf{Third}, existing solutions extend graph neural networks to social-aware recommendation has the potential to model social connections between users. However, incorporating the learned social- and knowledge-aware information still requires a tailored modeling to distill the high-order collaborative relations, with the joint preservation of local and global graph structural information.\\\vspace{-0.12in}

\noindent \textbf{The Present Work}. Having realized the importance of integrating heterogeneous user- and item-wise relations into the recommendation framework, as well as the corresponding challenges, this work develops \model\ (\underline{\textbf{S}}elf-Supervised \underline{\textbf{M}}etagraph \underline{\textbf{I}}nformax \underline{\textbf{N}}etworks), a knowledge-aware social recommendation architecture that simultaneously captures relational heterogeneity across users and items. Specifically, in order to handle relation heterogeneity, we propose a metapath-guided heterogeneous graph neural module to distill the multifaceted user-user and item-item relationships with the exploration of social network homophily, item knowledge graph dependencies as well as behavioral-level co-interactive patterns. \model\ first designs a meta relation-guided message passing architecture which maintains the type-specific relation representation space. Following the metepath-specific graph encoding layer, we further conduct the cross-metapath aggregation based on attention mechanism, to learn summarized representation vectors which capture the comprehensive semantics ingrained in the heterogeneous relations among users and items.

After encoding the relation heterogeneity (multifaceted social and knowledge graph dependencies) across users and items, we next incorporate the user-user and item-item relations into the encoding process of user-item interactive patterns, with a self-supervised graph mutual information learning architecture. Different from most of existing graph neural network-based collaborative filtering models which merely perform information propagation across local neighbors, we believe it is of critical importance to develop a relation encoder that investigates both local and global collaborative signals. Inspired by the effectiveness of introducing mutual dependence measurement in feature representation space~\cite{bachman2019learning,velickovic2019deep,peng2020graph}, \model\ transfers external knowledge from both user and item side into our interaction encoder under a self-supervised graph learning paradigm. Our \model\ performs the self-supervised data augmentation from two-folds: i) Inject the substructure awareness into the graph neural architecture to distill fine-grained semantics of local and global user-item interactive patterns; ii) Model the user-item high-order connectivity with the exploration of three different graph structure views: high-level substructure-aware global context, low-level node feature representation, and transformation characteristics of user-item interactive relations.

% Our \model\ is conceptually advantageous to those solutions by generalizing the mutual information learning paradigm into the recommendation scenario as supervised signals from two-folds.

To summarize, our work makes the following contributions:\vspace{-0.05in}

\begin{itemize}[leftmargin=*]

\item \textbf{General Aspects}. We emphasize the importance of jointly modeling social- and knowledge-aware relations from both user and item domains for better learning user's preference. We further incorporate the relation heterogeneity across users and items under a self-supervised learning paradigm, into the embedding space by exploring local and global collaborative similarities.\\\vspace{-0.1in}

\item \textbf{Methodologies}. To handle relation heterogeneity, we propose a metapath-guided heterogeneous graph neural network to maintain the user- and item-specific dependent representations through embedding propagation across graph layers. After that, an attentive cross-metapath aggregation layer is introduced to fuse the rich characteristics of multifaceted user- and item-wise inter-dependencies. In addition, \model\ allows the learned social- and knowledge-aware dependence to guide the user-item interaction embedding process from different structure views under a self-supervised graph learning architecture.\\\vspace{-0.12in}

\item \textbf{Experimental Findings}. We perform extensive experiments on three real-world datasets to demonstrate the superiority of our framework over state-of-the-art recommendation techniques.
% Through the model scalability study, we show that \model\ could achieve comparable efficiency as most advanced social-aware recommender systems.\vspace{-0.1in} %~\emph{https://github.com/anonymouscikm202105/\model}\vspace{-0.1in}

\end{itemize}

\section{Preliminary}
\label{sec:model}

\begin{figure*}[t]
    \centering
    %\vspace{-0.15in}
    \includegraphics[width=1\textwidth]{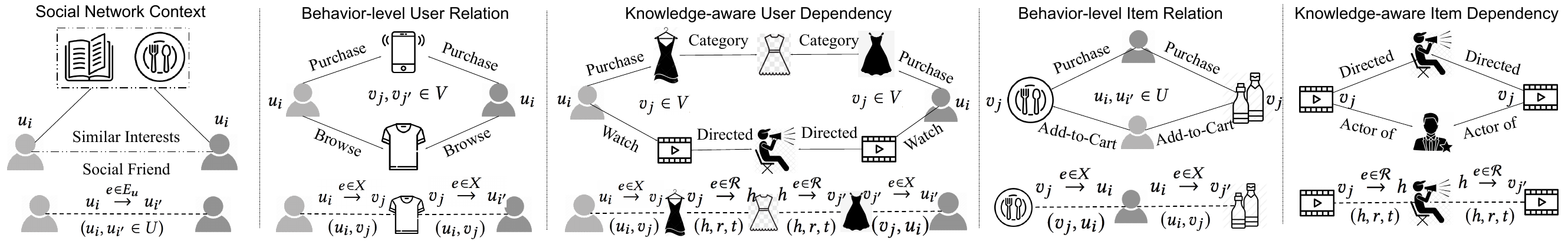}
    \vspace{-0.15in}
    \caption{Illustration of multifaceted meta-relations from both user and item domain.}
    \label{fig:meta_relation_examples}
    \vspace{-0.1in}
\end{figure*}

This section introduces key definitions and notations used in our work. Suppose we have $I$ users ($U=\{u_1,...,u_i,...,u_I\}$) and $J$ items ($V=\{v_1,...,v_j,...,v_J\}$) in a typical recommendation scenario, where $i$ and $j$ are identifier for user and item, respectively. We define the user-item interaction matrix $\textbf{X}=[x_{i,j}] \in \mathbb{R}^{I\times J}$, in which $x_{i,j}\in \{0,1\}$ indicates whether user $u_i$ has interacted with (\eg, purchased or clicked) item $v_j$, \ie, $x_{i,j}=1$ if interaction $(u_i,v_j)$ is observed and $x_{i,j}=0$ otherwise. In addition to the interaction data, we propose to consider side information from both users and items, and further present the relevant definitions which preserve the heterogeneous relationships across users and items as below.\\\vspace{-0.12in}

\noindent \textbf{Definition 2.1}. \textbf{User Social Graph} $G_u$. We define graph $G_u=\{U,E_u\}$ to denote social relations between users, where there exists an edge between user $u_i$ and $u_{i'}$ if they are socially connected.\\\vspace{-0.12in}

\noindent \textbf{Definition 2.2}.
\textbf{Item Relational Graph} $G_v$. To model the external knowledge from item side, we represent the item-wise relational structures in the form of dependency graph $G_v$. Specifically, the item inter-dependent signals are characterized with item-relation-item triples $(h,r,t)$, where $h, t\in \mathcal{V}$ and $r\in \mathcal{R}$. Here, $\mathcal{R}$ denotes the set of relations in $G_v$. For example, the relation $r$ can indicate that item $h$ belongs to the product category of $r$, or the co-interaction with the same user.\\\vspace{-0.12in}

% \textbf{Item Relational Graph} $G_v$. To model the external knowledge from item side, we represent the item-wise relational structures in the form of dependency graph $G_v$. Specifically, the item inter-dependent signals are characterized with item-relation-item triples $(v_j,r,v_{j'})$, where $v_j, v_{j'}\in \mathcal{V}$ and $r\in \mathcal{R}$. Here, $\mathcal{R}$ denotes the set of relationships between items. For example, the relation $r$ can indicate that item $h$ belongs to the product category of $t$, or the co-interaction with the same user.\\\vspace{-0.12in}

Based on the above notations, we define the collaborative heterogeneous graph $\mathcal{G}$ to integrate both user behavior as well side information from both user and item domains.\\\vspace{-0.12in}

\noindent \textbf{Definition 2.3}. \textbf{Collaborative Heterogeneous Graph} $\mathcal{G}$. Collaborative Heterogeneous Graph is defined as a unified graph $\mathcal{G}=\{\mathcal{V},\mathcal{E}\}$ which is associated with mapping functions for nodes: $\mathcal{V} \rightarrow \mathcal{A}$ and edges $\mathcal{E} \rightarrow \mathcal{B}$. Here, $\mathcal{A}$ and $\mathcal{B}$ denotes the sets of node and edge types, respectively, with $|\mathcal{A}|+|\mathcal{B}|>2$. In particular, the relation heterogeneity among users and items are modeled in $\mathcal{G}$ with the integration operations as: $\mathcal{V} = U \cup \{h,t\}\in G_v$ and $\mathcal{E} = E_u \cup \mathcal{R} \cup \{Interact\}$.\\\vspace{-0.12in}

\noindent \textbf{Task Formulation}. We now formulate our knowledge-aware social recommendation task as follows:\vspace{-0.05in}
\begin{itemize}[leftmargin=*]
\item \textbf{Input}: collaborative heterogeneous graph $\mathcal{G}=\{\mathcal{V},\mathcal{E}\}$ that collectively integrates user-item interaction $G_u$, user social graph $G_u$ and item relational graph $G_v$.
\item \textbf{Output}: the predictive function that infers the probability of an unknown behavior for target user $u_i$ interacting with item $v_j$.
\end{itemize}

\section{Methodology}
\label{sec:solution}

\begin{figure*}
	\centering
	\includegraphics[width=0.98\textwidth]{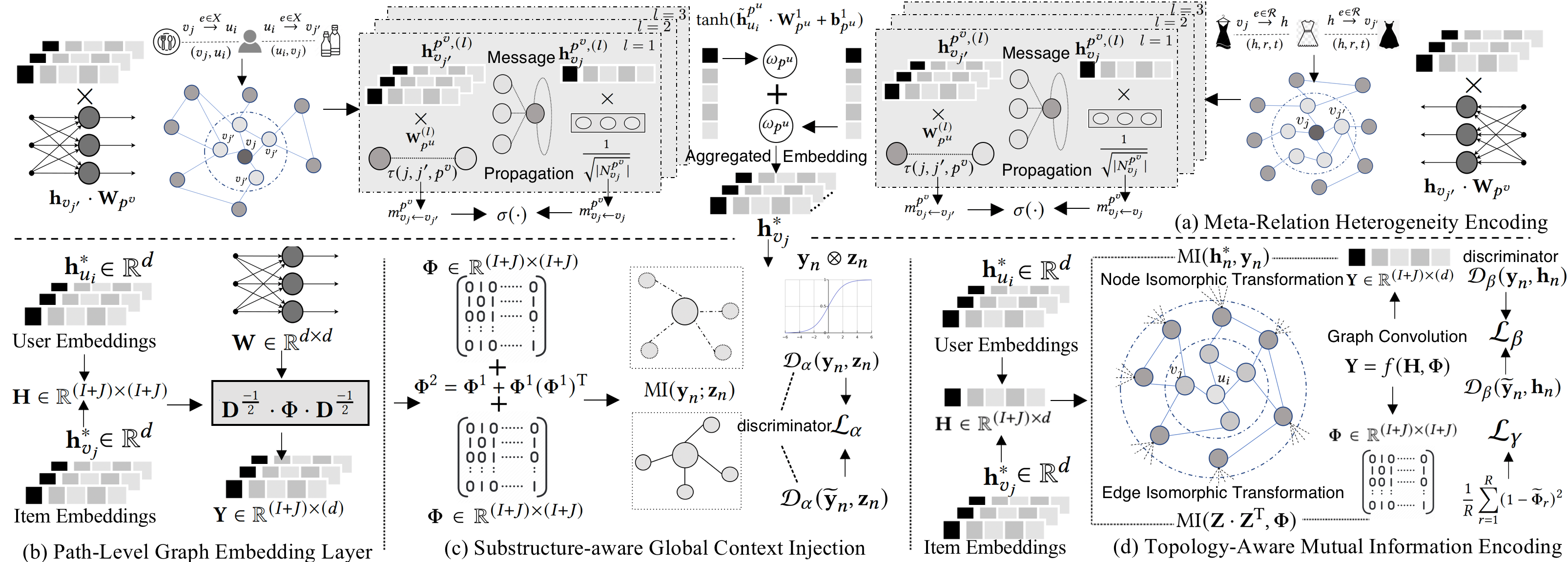}
	\vspace{-0.05in}
	\caption{(a). The integrative framework of metapath-guided heterogeneous neural network and cross-path aggregation network. (b-c) Substructure-aware global context modeling component that injects user-item high-order connectivity into the embedding space. (d) The topology-aware mutual information learning via capturing transformation property of user/item-specific characteristics and the underlying interactive patterns.}
	\label{fig:framework_1}
	\vspace{-0.05in}
\end{figure*}

This section presents the details of \model\ which consists of three key modules (shown in Figure~\ref{fig:framework_1}): i) Metapath-guided heterogeneous graph encoder which simultaneously captures the multifaceted social effects and item inter-dependent relations. ii) Cross-metapath aggregation network that integrates the semantic learned by metapath-specific representations. iii) Self-supervised graph mutual information learning which injects the social- and knowledge-aware embeddings into the user-item interaction modeling.

\subsection{Meta-Relation Heterogeneity Encoding}
We first develop a metapath-guided heterogeneous graph neural encoder to distill semantic information of user-wise relations and item-wise inter-dependencies, based on the metapath instances over the collaborative heterogeneous graph $\mathcal{G}$. In our work, a metapath instance $p$ represents a vertex sequence to capture the structural and semantic relation between objects~\cite{fang2019m,dong2017metapath2vec}. In particular, $p$ is formally defined in the form of $\mathcal{A}_1 \stackrel{\mathcal{E}_1}{\rightarrow} \mathcal{A}_2 \stackrel{\mathcal{E}_2}{\rightarrow} \cdots \stackrel{\mathcal{E}_C}{\rightarrow} \mathcal{A}_{C+1}$, which characterizes the meta-relations between vertex type $\mathcal{A}_1$ and $\mathcal{A}_{C+1}$ ($C$ denotes the length of metapath instance $p$).\\\vspace{-0.12in}

\noindent \textbf{Definition 3.1}. \textbf{Metapath-based Neighbor}. Based on the above present metapath instance $p$, nodes ($v' \in \mathcal{V}$) which are connected with $v$ through metapath instances in graph $\mathcal{G}$, are defined as metapath-based neighbors $N_v^p$ of $v$ under the metapath instance of $p$. Note that two nodes are connected through different metapath instances are regarded as different neighborhood relations.\\\vspace{-0.12in}

\subsubsection{\bf Multifaceted Meta-Relations Generation}
In our recommendation scenario, we extract different types of metapath instances for user and item domain, respectively, via exploring multifaceted social- and knowledge-aware dependence effects. Here, we define $p^u$ and $p^v$ to represent the extracted individual metapath instance for user and item, respectively. The corresponding instance sets are denoted as $P^u$ and $P^v$ ($p^u\in P^u$, $p^v\in P^v$). We present illustrated examples of multifaceted meta-relations from both user and item domain in Figure~\ref{fig:meta_relation_examples}.

\begin{itemize}[leftmargin=*]

\item \textbf{Social Network Context}. Social network homophily represents the static social relationships across users given their established social connections (\eg, friendship, community of interest). The generated user relation instance within the social network context is $u_i \stackrel{e\in E_u}{\rightarrow} u_{i'}$ ($u_i, u_{i'}\in U$)

\item \textbf{Behavior-level User Relation}. In addition to the static social network connections, we propose to model the social influence with the consideration of user behaviors. The corresponding metapath instance $p^u$ over graph $\mathcal{G}$ for behavior-level influence is $u_i \stackrel{e\in X}{\rightarrow} v_j \stackrel{e\in X}{\rightarrow} u_{i'}$ ($u_i, u_{i'}\in U$ and $v_j\in V$).

\item \textbf{Knowledge-aware User Dependency}. With the incorporation of item relational graph information into our recommendation framework, we further integrate the external knowledge graph signals with behavior-level user influence, so as to generate another type of metapath instance for user dependency, \ie, $u_i \stackrel{e\in X}{\rightarrow} v_j \stackrel{e\in \mathcal{R}}{\rightarrow} r \stackrel{e\in \mathcal{R}}{\rightarrow} v_{j'} \stackrel{e\in X}{\rightarrow} u_{i'}$.

\end{itemize}

For the item domain, we consider two types of metapaths:

\begin{itemize}[leftmargin=*]

\item \textbf{Behavior-level Item Relation}. We consider the influence between items with their attractiveness based on user interaction behavior. The attractiveness-based item metapath instance $p^v$ is generated as $v_j \stackrel{e\in X}{\rightarrow} u_i \stackrel{e\in X}{\rightarrow} v_{j'}$ ($v_j, v_{j'}\in V$ and $u_i\in U$).

\item \textbf{Knowledge-aware Item Dependency}. Since the relation graph $G_v$ characterizes semantic relatedness across items, we also model implicit item-wise dependencies with entity-based metapath instance of $v_j \stackrel{e\in \mathcal{R}}{\rightarrow} r \stackrel{e\in \mathcal{R}}{\rightarrow} v_{j'}$. $r$ denotes the intermediate entity connecting item $v_j$ and $v_{j'}$ with different relationships.

\end{itemize}

\subsubsection{\bf Metapath-guided Message Propagation}
After obtaining the metapath instances $p \in P$ with the preservation of multifaceted relational structures from both user and item side, we propose to encode the underlying relation heterogeneity with a graph-based neural architecture. In our metapath-guided message propagation framework, the embedding of individual node (\ie, user $u_i \in U$ and item $v_j \in V$) is generated with a graph neural layer aggregating feature information from metapath-based neighbors.

In particular, given the metapath instance of $p^u$ and $p^v$, we define the propagated message between ($u_i$,$u_{i'}$) and ($v_j$,$v_{j'}$) as $m^{p^u}_{u_i \leftarrow u_{i'}}$ and $m^{p^v}_{v_j \leftarrow v_{j'}}$, respectively. The corresponding embedding propagation process between users and items can be represented as follows:
\begin{align}
\label{eq:message_passing}
m^{p^u}_{u_i \leftarrow u_{i'}} = \psi(\textbf{h}^{p^u}_{u_{i'}};\tau(i,i',p^u));~m^{p^v}_{v_j \leftarrow v_{j'}} = \psi(\textbf{h}^{p^v}_{v_{j'}}, \tau(j,j',p^v)) 
\end{align}

Our graph-structured message passing paradigm for neighborhood feature aggregation at layer $l$ for node $u_i$ is formalized as:
\begin{small}
\begin{align}
\textbf{h}_{u_i}^{p^u,(l+1)} = \sigma(\textbf{h}^{p^u,(l)}_{u_i} \textbf{W}_{p^u}^{(l)} \frac{1}{|N_{u_i}^{p^u}|} + \sum_{u_{i'} \in N_{u_i}^{p^u}} \textbf{h}^{p^u,{(l)}}_{u_{i'}} \textbf{W}_{p^u}^{(l)} \frac{1}{\sqrt{|N_{u_i}^p||N_{u_{i'}}^p|}}) \nonumber
% \textbf{h}_{v_j}^{p^v,(l+1)} = \sigma(\textbf{h}^{p^v,(l)}_{v_j} \textbf{W}_{p^v}^{(l)} \frac{1}{\sqrt{|N_{v_j}^{p^v}|}} + \sum_{v_{j'} \in N_{v_j}^{p^v}} \textbf{h}^{p^v,{(l)}}_{v_{j'}} \textbf{W}_{p^v}^{(l)} \frac{1}{\sqrt{|N_{v_j}^{p^v}||N_{v_{j'}}^{p^v}|}}) \nonumber
\end{align}
\end{small}
\noindent The embedding propagation is performed through multiple ($L$) graph layers to update embeddings over metagraph neighborhoods with the $p$-th metapath instance. $\sigma(\cdot)$ indicates the PReLU function for dealing with feature non-linearities. We then apply the concatenate operation over the cross-layer social- and knowledge-aware node embeddings (\eg, $\textbf{h}_{u_i}^{p^u,(l)}$, $\textbf{h}_{v_j}^{p^v,(l)}$). $\textbf{W} \in \mathbb{R}^{d\times d}$ is the transformation matrix and $\textbf{Y} \in \mathbb{R}^{(I+J)\times (d)}$ denotes the user/item local-level embeddings encoded from their graph-based connectivity.

\subsection{Cross-metapath Aggregation Network}
We develop an aggregation layer, which is built upon the attention mechanism, to perform the cross-metapath information aggregation. The aggregation network aims to automatically learn a normalized importance weight $\omega_{p}$ for each metapath instance $p$, which is formally represented with the following operation. Here, we only present the cross-metapath aggregation layer for user-wise meta-relations. The same aggregation mechanism will be adopted for encoding relationship heterogeneity between items. 
\begin{align}
\textbf{q}_{u_i}^p = \text{tanh}(\tilde{\textbf{h}}_{u_i}^{p^u} \cdot \textbf{W}_{p^u}^1 + \textbf{b}_{p^u}^1);~~\hat{\omega}_{p^u} = \textbf{q}_{u_i}^{p} \cdot \textbf{W}_{p^u}^2 \nonumber\\
\omega_{p^u} = \frac{exp(\hat{\omega}_{p^u})}{\sum_{p^u\in P^u} exp(\hat{\omega}_{p^u})};~~\textbf{h}_{u_i}^{*} = \sum_{p^u\in P^u} \omega_{p^u} \cdot \tilde{\textbf{h}}_{u_i}^{p^u}
\label{eq:attention}
\end{align}
\noindent where $\textbf{W}_{p^u}^1 \in \mathbb{R}^{d\times d}$ and $\textbf{W}_{p^u}^2 \in \mathbb{R}^{d\times 1}$ are parameterized attention transformation matrices. $\textbf{b}_{p^u}^1 \in \mathbb{R}^{d}$ is the attentive bias term. The estimated importance weight $\omega_{p^u}$ reflects the importance of each type of meta-relation between users and items, \eg, social network homophily, behavior-level user influence and knowledge-aware item-wise relations. The summarized embeddings $\textbf{h}_{u_i}^{*}$, $\textbf{h}_{v_j}^{*}$ aggregate semantics across heterogeneous user-user and item-item meta-relations.

\subsection{Self-Supervised Mutual Information Learning}
After preserving and extracting the heterogeneous relational structures into the embedding space of users and items, we aim to inject high-order interactive patterns into our social recommendation model. Towards this end, we design a self-supervised mutual information learning architecture to facilitate learning expressive representations which jointly exhibit the local-level node-specific user/item characteristics and global-level graph dependencies. Motivated by the graph learning framework in~\cite{peng2020graph}, this component is built upon the self-supervised learning framework to construct supervising signals with auxiliary optimization objective.

\subsubsection{\bf Node-Level Graph Embedding Layer}
In our framework, we generate the feature representations $\textbf{H} \in \mathbb{R}^{(I+J)\times d}$ for graph nodes which consist of fused user/item embeddings ($\textbf{h}_{u_i}^{*}$, $\textbf{h}_{v_j}^{*}$), with the preservation of social- and knowledge-aware heterogeneous relations (encoded from our cross-metapath aggregation network). In addition, we define $\boldsymbol{\Phi} \in \mathbb{R}^{(I+J)\times (I+J)}$ matrix to represent the node adjacent relations over the user-item interaction graph, in which each entry $\phi=1$ if the corresponding user $u_i$ adopts item $v_j$ ($x_{i,j}=1$ in interaction matrix $\textbf{X}$) and $\phi=0$ otherwise. The encoding process is formally given as follows:
\begin{align}
\textbf{Y} = f(\textbf{H}, {\boldsymbol{\Phi}}) = \sigma(\textbf{D}^{\frac{-1}{2}} \cdot \boldsymbol{\Phi} \cdot \textbf{D}^{\frac{-1}{2}} \cdot \textbf{H} \cdot \textbf{W}) 
\end{align}
\noindent $\textbf{W} \in \mathbb{R}^{d\times d}$ is the transformation matrix , $\textbf{D} \in \mathbb{R}^{(I+J)\times (I+J)}$  is the degree matrix which contains information about the degree of each vertex and $\textbf{Y} \in \mathbb{R}^{(I+J)\times (d)}$ represents the user/item embeddings encoded from their graph-based connectivity.

\subsubsection{\bf Substructure-aware Global Context Injection}
It is worth mentioning that substructure often reflects unique characteristics and relation signals of graph data for network representation~\cite{wang2019adversarial,peng2020graph}, and collaborative user and item similarities could be captured sufficiently without constructing long-range graph-based connections across the entire user-item graph. Hence, we incorporate the substructure awareness into global context-enhanced user-item relation modeling from the low-level node and high-level graph topological information. In particular, we inject the substructure awareness into the user-item interaction graph by generating the $k$-order substructure-aware adjacent matrix based on the following element-wise addition operations across different order-specific relation adjacent matrices (\eg, $\boldsymbol{\Phi}^{1}$, $\boldsymbol{\Phi}^{(k-1)}$):
\begin{align}
\boldsymbol{\Phi}^{(k)} = \boldsymbol{\Phi}^{(k-1)} + \mathop{\underbrace{(\boldsymbol{\Phi}^1 \cdot (\boldsymbol{\Phi}^1)^\mathrm{T})\cdot \cdots \boldsymbol{\Phi}^1}}\limits_{k~\text{matrix instances}}
\end{align}
\noindent where $\boldsymbol{\Phi}^1$ represents the original adjacent matrix of user-item interaction graph by considering $1$-hop neighborhood relations as edges between nodes. $\boldsymbol{\Phi}^k$ incorporates all $k$-hop
neighbors into the adjacent relations. For example, $\boldsymbol{\Phi}^2 = \boldsymbol{\Phi}^1 + \boldsymbol{\Phi}^1 (\boldsymbol{\Phi}^1)^\mathrm{T}$ and $\boldsymbol{\Phi}^3 = \boldsymbol{\Phi}^1 + \boldsymbol{\Phi}^1 (\boldsymbol{\Phi}^1)^\mathrm{T} + (\boldsymbol{\Phi}^1 (\boldsymbol{\Phi}^1)^\mathrm{T}) \boldsymbol{\Phi}^1$. Given the $k$-order substructure-aware adjacent matrix $\boldsymbol{\Phi}^{(k)}$, we generate the graph-level representations:
\begin{align}
\textbf{Z} = \frac{\boldsymbol{\Phi}^{(k)}}{\hat{\boldsymbol{\Phi}}^{(k)}} \cdot \textbf{Y} ;~~
\hat{\boldsymbol{\Phi}}^{(k)}_n = \sum_{r'=1}^{(I+J)} \boldsymbol{\Phi}^{(k)}_{r,r'};~~\boldsymbol{\Phi}^{(k)}_n \in \mathbb{R}^{(I+J)}
\end{align}
\noindent where $\textbf{Z} \in \mathbb{R}^{(I+J)\times d}$ represents the learned substructure-aware global embeddings and each row of which represents the substructure-aware global embedding corresponding to each node (user $u_i$ or item $v_j$) in the interaction graph. It injects the global graph-structured context into latent representations $\textbf{Y}$. $\hat{\boldsymbol{\Phi}}^{(k)} \in \mathbb{R}^{(I+J)}$ is the degree matrix which contains the node degree of individual node. Next, we propose to capture both local and global user-item interactive patterns via encoding the mutual information between node embedding $\textbf{y}_n \in \textbf{Z}$ and the corresponding high-level graph representation $\textbf{z}_n \in \textbf{Z}$.
Following the paradigm in~\cite{hjelm2018learning}, we define our global context-aware discriminator $\mathcal{D}_{\alpha}(\textbf{y}_n,\textbf{z}_n) = \delta(\textbf{y}_n \otimes \textbf{z}_n)$ ($\delta(\cdot)$ denotes the sigmoid function) to classify the pairwise relationship between $\textbf{y}_n$ and $\textbf{z}_n$ from the joint distribution with the aim of maximizing the mutual information $\text{MI}(\textbf{y}_n;\textbf{z}_n)$. In particular, the embedding pair $(\textbf{y}_n, \textbf{z}_n)$ is fed into our discriminator as positive instance. We generate negative instance $(\widetilde{\textbf{y}}_n, \textbf{z}_n)$. We further define the objective function with the modeling of global context-aware mutual information $\text{MI}(\textbf{y}_n;\textbf{z}_n)$ as follows:
\begin{align}
\mathcal{L}_{\alpha} = \frac{-1}{I+J} (\sum_{n=1}^{I+J} log \mathcal{D}_{\alpha}(\textbf{y}_n,\textbf{z}_n) + \sum_{n=1}^{I+J} log [ 1 - \mathcal{D}_{\alpha}(\widetilde{\textbf{y}}_n,\textbf{z}_n)])
\end{align}
\noindent Minimizing the above cross-entropy-based loss contributes to the mutual information maximization and capturing the user-item high-order collaborative relations from locally to globally.

\subsubsection{\bf Topology-Aware Mutual Information Encoding}
We propose to consider the properties of the topological information over the user-item interaction graph (motivated by the graph structure modeling in~\cite{morris2019weisfeiler,peng2020graph}). Specifically, we define our topology-aware mutual information as $\text{MI}(\textbf{h}_n^*, \textbf{y}_n)$ and $\text{MI}(\textbf{Z} \cdot \textbf{Z}^\mathrm{T}, \boldsymbol{\Phi})$ which reflects the transformation property of node and edge structural feature, respectively. To achieve this goal, we design our discriminator $\mathcal{D}_{\beta}(\textbf{y}_n, \textbf{h}_n) = \delta(\textbf{y}_n \otimes \textbf{h}_n)$, to endow our graph embedding layer $f(\textbf{H}, {\boldsymbol{\Phi}})$ with the ability of preserving characteristics of node dimension between $\textbf{y}_n$ and $\textbf{h}_n$. The corresponding optimized objective function is given:
\begin{align}
\mathcal{L}_{\beta} = \frac{-1}{I+J} (\sum_{n=1}^{I+J} log \mathcal{D}_{\beta}(\textbf{y}_n,\textbf{h}_n) + \sum_{n=1}^{I+J} log [ 1 - \mathcal{D}_{\beta}(\widetilde{\textbf{y}}_n,\textbf{h}_n)])
\end{align}
% \noindent The corresponding positive and negative instances are generated with 

Additionally, another dimension of transformation lies in the encoded user-item interactions (relational edges) from our graph embedding layer. The loss with edge transformation is defined as:
% The re-constructed connections $\widetilde{\boldsymbol{\Phi}}$ between user and item are generated as: $\widetilde{\boldsymbol{\Phi}} = \textbf{Z} \cdot \textbf{Z}^\mathrm{T}$ ($\textbf{Z} \in \mathbb{R}^{(I+J)\times d}$).
\begin{align}
\mathcal{L}_{\gamma} = \frac{1}{R} \sum_{r=1}^R (1 - \widetilde{\boldsymbol{\Phi}}_{r})^2;~~ \widetilde{\boldsymbol{\Phi}} = \textbf{Z} \cdot \textbf{Z}^\mathrm{T}~~(\textbf{Z} \in \mathbb{R}^{(I+J)\times d})
\end{align}
\noindent $\widetilde{\boldsymbol{\Phi}}$ indicates the re-constructed user-item connections which are generated with our encoded user and item representations $\textbf{z}_{u_i}$ and $\textbf{z}_{v_j}$. $R$ (indexed by $r$) denotes the number of non-zero elements in $\widetilde{\boldsymbol{\Phi}}$. With the preservation of the transformation of both nodes and their connections, we augment the representation learning by incorporating topology-aware user-item interactions.

\subsection{The Learning Process of \model}

\subsubsection{\bf Optimization Objective}
We define our joint optimized objection $\mathcal{L}$ with the integration of heterogeneous relationship learning across users and items, and user-item interaction modeling via graph-structured mutual information encoder, based on Bayesian personalized ranking loss:
\begin{align}
\mathcal{L} = \frac{-1}{N} \Big ( \sum_{u_i,v_j^+,v_j^- \in O} log [\delta(\textbf{h}_{u_i}^*, \textbf{h}_{v_j^+}^*) - \delta(\textbf{h}_{u_i}^*, \textbf{h}_{v_j^-}^*)] \nonumber\\
+ \lambda_0 \mathbin\Vert \Theta \mathbin\Vert^2 +  \lambda_{\alpha} \cdot \mathcal{L}_{\alpha} + \lambda_{\beta} \cdot \mathcal{L}_{\beta} +  \lambda_{\gamma} \cdot \mathcal{L}_{\gamma} \Big )
\end{align}
\noindent where $O$ is the training instance consisting of positive ($(u_i,v_j^+)$) and negative ($(u_i,v_j^-)$) samples corresponding to the observed and unobserved interactions between user and item. $\Theta$ denotes the trainable parameters for regularization with the strength of $\lambda_0$. $\lambda_{\alpha}$, $\lambda_{\beta}$ and $\lambda_{\gamma}$ balances the loss functions with mutual information maximization from different views.

\subsubsection{\bf Model Complexity Analysis}
In this subsection, we perform time complexity analysis of our \model\ method. Particularly, the cost of cross-layer message propagation in our metapath-guided heterogeneous graph neural network lies in the $O(|E|\times L\times d)$, where $|E|$ denotes the number of meta-connections, $L$ represents the depth of our graph neural network, and $d$ denotes the latent dimensionality.
% In our multi-view mutual information learning component, the complexity of global context injection is $O(|E_k|\times d + (I+J)\times d)$, where $|E_k|$ denotes the number of relational edges over the user-item interaction graph with the $k$-hop graph substructure awareness. Moreover, the topology-aware mutual information encoding process has computational cost of $O(|E_{u-v}|\times d)$ and $O((I+J)\times d)$ for capturing isomorphic transformation property of nodes and edges, respectively. Here, $|E_{u-v}|$ indicates the number of non-zero elements in interaction matrix $\textbf{X}$. 
The final prediction layer has the complexity of $O(|E_{u-v}|\times L\times d)$ . Given the component-specific complexity analysis, the overall computational complexity of \model\ is $O((|E|+|E_{u-v}|)\times L\times d)$, which is comparable to most of existing social recommendation approaches.% (as demonstrated in Section~\ref{sec:scalability}).

\section{Evaluation}
\label{sec:eval}

We evaluate the performance of our \emph{\model} recommendation framework on three public datasets collected from real-life platforms. This section aims to study the following research questions (RQs):

\begin{itemize}[leftmargin=*]

\item \textbf{RQ1}: How does \emph{\model} perform compared with various baselines?\\\vspace{-0.12in}

\item \textbf{RQ2}: How do different components (\eg, metapath-guided heterogeneous graph neural network, mutual information-augmented learning paradigm) affect the performance of \emph{\model}?\\\vspace{-0.12in}

\item \textbf{RQ3}: What is the impact of different types of meta-relations across users and items in the recommendation performance?\\\vspace{-0.12in}

% \item \textbf{RQ4}: How is recommendation accuracy of \emph{\model} when competing with baselines \wrt\ different interaction density degrees?\\\vspace{-0.12in}

\item \textbf{RQ4}: How do different settings of key hyperparameters affect the model prediction performance?\\\vspace{-0.12in}

\item \textbf{RQ5}: How do the learned latent representations benefit from the collectively encoding of social- and knowledge-enhanced user-item interactive patterns? \\\vspace{-0.12in}

% \item \textbf{RQ7}: How is the model scalability of our \emph{\model} framework?

\end{itemize}

\begin{table}[t!]
% \vspace{-0.1in}
\centering
\small
% \scriptsize
% \footnotesize
%\scriptsize
\caption{Statistical information of the datasets.}
\vspace{-0.15 in}
% \begin{adjustbox}{max width=.95\linewidth}
\begin{tabular}{l| c| c| c}
\hline
Dataset & Ciao & Epinions & Yelp \\
\hline
\# of Users & 6,776 & 15,210 & 161,305 \\
\# of Items & 101,415 & 233,929 & 114,852 \\
\# of Interactions & 271,573 & 644,715 & 1,118,645 \\
Interaction Density & 0.0395\% & 0.0181\% & 0.0060\% \\
%\# of User Meta-Relations & 23,417,128 & 27,574,212 & 70,074,693 & 32,268,828 \\
%Social Tie Density Degree & & 0.1806\% & 0.0296\% & 0.0121\%\\
%\# of Item Meta-Relations & 23,696,228 & 19,017,630 & 24,213,384 & 11,845,101\\
\hline
\end{tabular}
\vspace{-0.1in}
% \end{adjustbox}
\label{tab:data}
\end{table}

% \begin{table}[t!]
% % \vspace{-0.1in}
% \centering
% %\small
% % \scriptsize
% \footnotesize
% %\scriptsize
% \caption{Statistics of Experimented Datasets.}
% \vspace{-0.15 in}
% % \begin{adjustbox}{max width=.95\linewidth}
% \begin{tabular}{l| c| c| c| c}
% \hline
% Dataset & Ciao & Epinions & Yelp & E-commerce\\
% \hline
% \# of Users & 6,776 & 15,210 & 161,305 & 377,882 \\
% \# of Items & 101,415 & 233,929 & 114,852 & 211,768\\
% \# of Interactions & 271,573 & 644,715 & 1,118,645 & 2,499,673\\
% Interaction Density & 0.0395\% & 0.0181\% & 0.0060\% & 0.0031\%\\
% %\# of User Meta-Relations & 23,417,128 & 27,574,212 & 70,074,693 & 32,268,828 \\
% %Social Tie Density Degree & & 0.1806\% & 0.0296\% & 0.0121\%\\
% %\# of Item Meta-Relations & 23,696,228 & 19,017,630 & 24,213,384 & 11,845,101\\
% \hline
% \end{tabular}
% \vspace{-0.1in}
% % \end{adjustbox}
% \label{tab:data}
% \end{table}

\subsection{Experimental Settings}

\subsubsection{\bf Data Description.} Our \emph{\model} framework is evaluated on three real-world datasets. We present the data statistics in Table~\ref{tab:data} and elaborate details of individual dataset as below:

\begin{itemize}[leftmargin=*]
\item \textbf{Ciao and Epinions Data}\footnote{https://www.cse.msu.edu/~tangjili/datasetcode/truststudy.htm}. These two datasets serve as benchmarks for evaluating social recommender systems. Ciao and Epinions are two popular consumer review platforms in which users are free to establish their social connections.

\item \textbf{Yelp Data}\footnote{https://www.yelp.com/dataset/download}. This data is collected from Yelp platform and contains users' online friendships with respect to their similar interests. In Yelp, customers could provide feedback on local venues from different business categories.

% \item \textbf{E-Commerce Data}. We also perform evaluation of our framework on a dataset from a major commercial e-commerce platform. This data is constructed with users' online relationships and user-product clicking records.

\end{itemize}

% In our experiments, in addition to user-user dependencies and user-item interactions, we further generate the knowledge-aware item relations by extracting external knowledge (\eg, product categories, venue's business genres and locations) from item domain, which follows the similar data pre-processing strategies in knowledge-aware recommender systems~\cite{xin2019relational,wang2019kgat}.

\subsubsection{\bf Evaluation Protocols.}
In our evaluation, we adopt the leave-one-out strategy for training and testing set generation, which has been widely used in recommendation applications~\cite{qin2020sequential,wu2019neural}. To follow the similar experimental settings in~\cite{chen2019social,sun2019bert4rec} and enable the result evaluation in an efficient way, we pair the positive sample of the target user with 99 negative instances which have no interaction with this user. Since this work focus on top-N item recommendation scenario, we leverage two representative recommendation evaluation metrics: Hit Ratio (HR) and Normalized Discounted Cumulative Gain(NDCG). In our experiments, in addition to user-user dependencies and user-item interactions, we further generate the knowledge-aware item relations by extracting external knowledge (\ie, categories of item) from item domain.

\begin{table*}[t]
%\scriptsize
\footnotesize
%\small
%\centering
%\vspace{-0.10in}
\caption{Recommendation performance of different methods in terms of \emph{HR@10} and \emph{NDCG@10}.} 
\vspace{-0.1in}
\begin{center}
\setlength{\tabcolsep}{1.0mm}
\begin{tabular}{ c || c| c | c | c | c | c | c | c | c | c | c | c | c | c | c || c}
\hline
Dataset & Metrics & ~~PMF~~ & ~TrustMF~ & ~~DiffNet~ & ~SAMN~ & ~DGRec~ & EATNN & NGCF$_+$ & ~KGAT~ & ~~MKR~~ & GraphRec & DANSER & HERec & MCRec & HAN & ~\emph{\model}~ \\
\hline
\multirow{2}{*}{Ciao} & HR & 0.6385 & 0.6560 & 0.6747 & 0.6576 & 0.6653 & 0.6738 & 0.6945 & 0.6601 & 0.6793 & 0.6825 & 0.6730 & 0.6800 & 0.6772 & 0.6589 & \textbf{0.7108} 	\\
\cline{2-17}
& NDCG  & 0.4420 & 0.4532 & 0.4636 & 0.4561 & 0.4593 & 0.4665 & 0.4894 & 0.4512 & 0.4589 & 0.4730 & 0.4521 & 0.4712 & 0.4708 & 0.4469 & \textbf{0.5012} \\
\cline{1-17}
\multirow{2}{*}{Epinions} & HR & 0.7445 & 0.7502 & 0.7699 & 0.7592 & 0.7603 & 0.7650 & 0.7984 & 0.7510 & 0.7647 & 0.7723 & 0.7714 & 0.7642 & 0.7630 & 0.7505 & \textbf{0.8179}	\\
\cline{2-17}
& NDCG  & 0.5491 & 0.5551 & 0.5702 & 0.5614 & 0.5668 & 0.5663 & 0.5945 & 0.5578 & 0.5669 & 0.5751 & 0.5741 & 0.5495 & 0.5326 & 0.5275 & \textbf{0.6137}	\\
\hline
\multirow{2}{*}{Yelp} & HR & 0.7554 & 0.7791 & 0.8048 & 0.7910 & 0.7950 & 0.8031 & 0.8265 & 0.7881 & 0.8005 & 0.8098 & 0.8077 & 0.7928 & 0.7869 & 0.7731 & \textbf{0.8478} \\
\cline{2-17} 
& NDCG & 0.5165 & 0.5424 & 0.5670 & 0.5516 & 0.5593 & 0.5560 & 0.5854 & 0.5501 & 0.5635 & 0.5679 & 0.5692 & 0.5612 & 0.5590 & 0.5604 & \textbf{0.5993} \\
% \cline{1-17}
% \multirow{2}{*}{E-Com} & HR & 0.6521  & 0.6796 & 0.7020 & 0.6880 & 0.6903 & 0.7028 & 0.7373 & 0.6814 & 0.7034 & 0.7105 & 0.7168 & 0.6973 & 0.6902 & 0.6841 & \textbf{0.7575} \\
% \cline{2-17}
% & NDCG  & 0.4523 & 0.4710 & 0.4990 & 0.4796 & 0.4911 & 0.5087 & 0.5261 & 0.4861 & 0.4934 & 0.5111 & 0.5213 & 0.4736 & 0.4551 & 0.4390 & \textbf{0.5468}\\
\cline{1-17}
\hline
\end{tabular}
\end{center}
\vspace{-0.05in}
\label{tab:result}
\end{table*}

\subsubsection{\bf Baselines for Comparison}
We compare \emph{\model} with various state-of-the-art baselines as shown below:

\begin{itemize}[leftmargin=*]

\item \textbf{PMF}~\cite{mnih2008probabilistic}: it is a probabilistic matrix factorization method to factorize users and items into latent feature vectors. \\\vspace{-0.12in}

\item \textbf{TrustMF}~\cite{yang2016social}: The proposed new matrix factorization-based framework incorporates the users' trust relationships into the embedding process, to augment the user preference modeling.\\\vspace{-0.12in}

\item \textbf{EATNN}~\cite{chen2019efficient}: It is an adaptive transfer learning model with attention network to capture the interplay between users and items.\\\vspace{-0.12in}

\item \textbf{SAMN}~\cite{chen2019social}: A dual-stage attention model is developed to approximate the user-wise relations with their social neighbors.\\\vspace{-0.12in}

\item \textbf{DiffNet}~\cite{wu2019neural}: It recursively updates users' embeddings with a influence diffusion component. The user-item collaborative signals are captured through a fusion layer.\\\vspace{-0.12in}% The number of influence diffusion layers are set as 2 to achieve the best performance.\\\vspace{-0.1in}

\item \textbf{GraphRec}~\cite{fan2019graph}: In this framework, a graph attention framework is developed for embedding propagation between users for social relation aggregation.\\\vspace{-0.12in}
%Three hidden layers are utilized for neural components.

\item \textbf{DANSER}~\cite{wu2019dual}: It designs a dual-stage graph attention network to model multifaceted social effects in recommendation.\\\vspace{-0.12in}

\item \textbf{DGRec}~\cite{song2019session}: This social recommender system attempts to model dynamic user behavioral patterns and social influence by integrating the recurrent neural network with graph attention layer. \\\vspace{-0.12in}

\item \textbf{NGCF$_+$}~\cite{wang2019neural}: It is a state-of-the-art graph neural network-based collaborative filtering model. We extend it to perform message propagation over our heterogeneous graph $\mathcal{G}$. \\\vspace{-0.12in}

\item \textbf{KGAT}~\cite{wang2019kgat}: It incorporates the knowledge graph information into the collaborative relation modeling and performs recursively embedding propagation between connected node instances.\\\vspace{-0.12in}
% Three graph neural layers are stacked as reported in their released code.

\item \textbf{MKR}~\cite{wang2019multi}: MKR is a knowledge-aware recommender system based on multi-task learning paradigm, which leverages knowledge graph representation task to assist the encoding of user-item interactions.\\\vspace{-0.12in}

\item \textbf{HERec}~\cite{shi2018heterogeneous}: it is a heterogeneous network embedding approach with meta-path random walk for node sequence generation.\\\vspace{-0.12in}

\item \textbf{MCRec}~\cite{hu2018leveraging}: this recommendation method incorporates the meta-path based context into the neural co-attention mechanism for recommendation performance improvement.\\\vspace{-0.12in}

\item \textbf{HAN}~\cite{wang2019heterogeneous}: it captures the heterogeneity of graph using the attentive encoder to differentiate relations between users and items.

\end{itemize}

\subsubsection{\bf Parameter Settings}
% We present the hyperparameter configuration details of our \emph{\model} as well as the settings of baselines.\\\vspace{-0.12in}

We implement \emph{\model} in PyTorch and optimize it with Adam optimizer. The learning process is performed with the batch size tuned amongst [1024, 2048, 4096, 8192] and learning rate of $5e^{-2}$ (with decay rate of 0.95 every epoch). $L_2$ regularization term with $\lambda_0$ of $0.05$ is adopted over $\Theta$. The number of hidden state dimensionality $d$ is tuned from [8,16,32,64,128]. The depth $L$ of our graph neural layers in our relation heterogeneity encoder is chosen from 1 to 3. The number of neighbors $k$ for global-level context injection in our graph mutual information learning architecture is set as 2. The sensitivities of those key hyperparameters have been investigated in Section~\ref{sec:parameter}. The performance of most neural methods is evaluated based on their source code.

% To ensure a fair comparison, we keep the settings of each baseline approach to be consistent with the reports in the paper. Furthermore, the grid-search strategy is adopted for hyperparameter tuning to achieve the optimal performance.

%The source code and experimented datasets will be released after the acceptance.

\begin{table}[t]
    \vspace{-0.05in}
	\caption{Ranking performance on Epinions dataset with varying Top-\textit{N} value in terms of \textit{HR@N} and \textit{NDCG@N}}
	\vspace{-0.1in}
	\centering
    %\scriptsize
    %\small
    \footnotesize
	\setlength{\tabcolsep}{1mm}
	\begin{tabular}{|c|c|c|c|c|c|c|}
		\hline
% 		\multirow{2}{*}{Model}&\multicolumn{2}{c|}{@5}&\multicolumn{2}{c|}{@10}&\multicolumn{2}{c||}{@15}&\multicolumn{2}{c|}{@5}&\multicolumn{2}{c|}{@10}&\multicolumn{2}{c}{@15}\\
		%\cline{2-13}
		Model &HR@5 &NDCG@5 & HR@10 & NDCG@10 & HR@15 & NDCG@15\\
		\hline
        SAMN    & 0.6492 & 0.4851 & 0.7592 & 0.5614 & 0.7975 & 0.5492\\
        \hline
        EATNN   & 0.6642 & 0.5086 & 0.7650 & 0.5663 & 0.7646 & 0.5491\\
        \hline
        DGRec   & 0.6564 & 0.4970 & 0.7603 & 0.5668 & 0.7947 & 0.5467\\
        \hline
        DiffNet & 0.6428 & 0.4819 & 0.7699 & 0.5702 & 0.8009 & 0.5437\\
        \hline
        GraphRec & 0.6615 & 0.4941 & 0.7647 & 0.5669 & 0.7900 & 0.5414\\
        \hline
        DANSER & 0.6704 & 0.5297 & 0.7723 & 0.5751 & 0.8106 & 0.5562\\
        \hline
        NGCF$_+$  & 0.7002 & 0.5619 & 0.7984 & 0.5945 & 0.8419 & 0.6014\\
        \hline
        KGAT    & 0.6538 & 0.4952 & 0.7510 & 0.5578 & 0.7882 & 0.5403\\
        \hline
        MKR     & 0.6518 & 0.5056 & 0.7714 & 0.5741 & 0.8257 & 0.5712\\
        \hline
        \hline
        \emph{\model} & \textbf{0.7245}  & \textbf{0.5871} & \textbf{0.8179} & \textbf{0.6137} & \textbf{0.8622} & \textbf{0.6241}\\
        \hline
	\end{tabular}
	\label{tab:vary_k}
	\vspace{-0.1in}
\end{table}

\subsection{Performance Comparison (RQ1)}
Table~\ref{tab:result} summarizes the performance comparison between our developed \emph{\model} and competing state-of-the-arts on three datasets. We summarize the following observations:

\noindent (1). In terms of both metrics, we can observe that \emph{\model} consistently outperforms other alternatives in all cases, which suggests the effectiveness of our method by modeling heterogeneous social and item graph dependencies (based on user- and item-wise meta relation schema) under a self-supervised graph neural architecture.\\\vspace{-0.12in}

\noindent (2). Among various baselines, graph neural network social recommendation models (\eg, GraphRec and DANSER) perform better than others, which suggests that the graph-structured embedding aggregation paradigm is an effective solution for social-aware recommender systems. The consistent performance gap between our \emph{\model} and those GNN-based approaches, implies that encoding relation heterogeneity from both social and item domain could enhance graph neural architecture for recommendation. We further optimize those graph models by stacking more graph layers to fit the high-order user-item interaction, but have not observed clear improvement with $L \geq 3$.\\\vspace{-0.12in}

\noindent (3). While NGCF$_+$ also considers the cross-user and item dependencies under a graph learning architecture, by comparing with \emph{\model} and NGCF$_+$, we can observe that our develop mutual information learning framework results in further performance improvement compared with other graph-based model for capturing high-order user-item collaborative relations. \\\vspace{-0.12in}

\noindent (4). We can notice that recently proposed knowledge-aware recommendation techniques (\ie, KGAT and MKR) experience performance degeneration, which sheds light on the limitation of failing to comprehensively transfer the knowledge graph semantics into the users' behavior modeling. Different from them, \emph{\model} achieves better performance by utilizing the multifaceted relationships across users and items as self-supervised signals to guide the user preferences representation. \\\vspace{-0.12in}

\noindent (5). From the presented evaluation results in Table~\ref{tab:vary_k}, the performance superiority of our \emph{\model} can be observed under different top-$N$ recommended items. This observation further validates the effectiveness of our new recommendation framework.

% As shown in Table~\ref{tab:vary_k}, we notice the performance gain achieved by \emph{\model} over other competitors with different ranked top-$N$ positions, which further justifies the superior ranking performance. The recommendation accuracy improves with larger $N$ values.

%\vspace{-0.1in}
\subsection{Ablation Study of \emph{\model} Framework (RQ2)}
We conduct experiments to demonstrate the rationality of sub-components in our \emph{\model} model. To achieve this goal, we implement five simplified variants to show the influence of each model design.
\begin{itemize}[leftmargin=*]
\item \textbf{Effect of Relation Heterogeneity Modeling}. \emph{\model}-h. We remove the metapath-guided heterogeneous graph neural network to fuse multifaceted user- and item-wise relationships.\\\vspace{-0.1in}
\item \textbf{Effect of Self-Supervised Information Encoder}. \emph{\model}-s. We do not include the multi-view graph mutual information encoder into our framework. Instead, we merely utilize graph convolution network to perform embedding propagation over the collaborative heterogeneous graph $\mathcal{G}$.\\\vspace{-0.1in}

\item \textbf{Effect of Global Context Injection}. \emph{\model}-g. We remove the mutual information $\text{MI}(\textbf{y}_n;\textbf{z}_n)$ to inject the global-level collaborative signals based on graph substructure.\\\vspace{-0.1in}

\item \textbf{Effect of Topology-aware Isomorphic Transformation}. \emph{\model}-t. In this variant, we remove the constrains of the isomorphic transformation in terms of topological node and edge characteristics, \ie, $\text{MI}(\textbf{h}_n^*, \textbf{y}_n)$ and $\text{MI}(\textbf{Z} \cdot \textbf{Z}^\mathrm{T}, \boldsymbol{\Phi})$.\\\vspace{-0.1in}

\item \textbf{Effect of Cross-Metapath Aggregation}. \emph{\model}-a. We replace the cross-metapath aggregation layer with the mean pooling operation without differentiating the importance of metapath-specific relational feature representations.

\end{itemize}

The study results are shown in Table~\ref{tab:ablation_result}. We summarize the following findings. (1) The performance is improved with our metapath-guided self-supervised heterogeneous graph network for capturing diversified social and item-wise relations. (2) The performance gap between \emph{\model} and \emph{\model}-s verifies the rationality of considering multi-view mutual relational signals in recommendation under a self-supervised learning paradigm. (3) Injecting the substructure-aware graph global context imposes positive effect for encoding high-order user-item connectivity. (4) The careful consideration of relational learning gains more insights for modeling graph-based user-item interaction topological structure. (5) \emph{\model} outperforms \emph{\model}-a demonstrates the efficacy of encoding the importance of each metapath instance in an explicit way for generating user and item embeddings. This is because that each type of user-user and item-item relations contribute differently for the learning of user-item interactive patterns.

\begin{table}[t]
%\scriptsize
%\footnotesize
\small
%\centering
%\vspace{-0.10in}
\caption{Ablation Study of \emph{\model} Framework.} 
\vspace{-0.15in}
\begin{center}
\setlength{\tabcolsep}{1.0mm}
\begin{tabular}{ c || c| c | c | c | c | c }
\hline
Variants & \multicolumn{2}{c|}{Ciao} & \multicolumn{2}{c|}{Epinions} & \multicolumn{2}{c}{Yelp} \\
\hline
Metrics & HR & NDCG & HR & NDCG & HR & NDCG \\
\hline
\hline
\emph{\model}-h &0.6864 &0.4706 & 0.8024 & 0.5934 & 0.8362 & 0.5887  \\
\emph{\model}-s &0.6800 &0.4669 & 0.8034 & 0.5892& 0.8422 & 0.5907   \\
\emph{\model}-g &0.6913 &0.4769 & 0.8155&0.6114 & 0.8468&0.5903  \\
\emph{\model}-t &0.6833 &0.4743 & 0.8144&0.6069 &0.8456 &0.5982  \\
\emph{\model}-a &0.6838 &0.4645 & 0.8094&0.5909 &0.8425 &0.5894  \\
\hline
\emph{\model} &\textbf{0.6919} &\textbf{0.4751} & \textbf{0.8179} & \textbf{0.6137} & \textbf{0.8465} & \textbf{0.5989} \\
\hline
\end{tabular}
\end{center}
\vspace{-0.1in}
\label{tab:ablation_result}
\end{table}

% \begin{table}[t]
% %\scriptsize
% \footnotesize
% %\small
% %\centering
% %\vspace{-0.10in}
% \caption{Ablation Study of \emph{\model} Framework.} 
% \vspace{-0.1in}
% \begin{center}
% \setlength{\tabcolsep}{1.0mm}
% \begin{tabular}{ c || c| c | c | c | c | c | c | c }
% \hline
% Variants & \multicolumn{2}{c|}{Ciao} & \multicolumn{2}{c|}{Epinions} & \multicolumn{2}{c|}{Yelp} & \multicolumn{2}{c}{E-Commerce} \\
% \hline
% Metrics & HR & NDCG & HR & NDCG & HR & NDCG & HR & NDCG\\
% \hline
% \hline
% \emph{\model}-h &0.6864 &0.4706 & 0.8024 & 0.5934 & 0.8362 & 0.5887 & 0.7484 & 0.5371 \\
% \emph{\model}-s &0.6800 &0.4669 & 0.8034 & 0.5892& 0.8422 & 0.5907 & 0.7491 & 0.5375 \\
% \emph{\model}-g &0.6913 &0.4769 & 0.8155&0.6114 & 0.8468&0.5903 &0.7498 &0.5386 \\
% \emph{\model}-t &0.6833 &0.4743 & 0.8144&0.6069 &0.8456 &0.5982 &0.7564 &0.5403 \\
% \emph{\model}-a &0.6838 &0.4645 & 0.8094&0.5909 &0.8425 &0.5894 &0.7527 &0.5436 \\
% \hline
% \emph{\model} &\textbf{0.6919} &\textbf{0.4751} & \textbf{0.8179} & \textbf{0.6137} & \textbf{0.8465} & \textbf{0.5989} & \textbf{0.7575} & \textbf{0.5468} \\
% \hline
% \end{tabular}
% \end{center}
% \vspace{-0.1in}
% \label{tab:ablation_result}
% \end{table}

%\vspace{-0.1in}
\subsection{Effect Analyses of Meta-Relations (RQ3)}
To study the effect of multifaceted meta-relations incorporated in our metapath-guided heterogeneous graph neural network, we also perform ablation studies with the integration of different meta-relations from user and item domain. In specific, we generate four contrast variants of \emph{\model} as below for performance comparison:
\begin{itemize}[leftmargin=*]
\item \emph{\model}-UU: \emph{\model} without the user-user online social ties to capture the social network homophily, \ie, $u_i \stackrel{e\in E_u}{\rightarrow} u_{i'}$.\\\vspace{-0.1in}

\item \emph{\model}-UIU\&IUI: \emph{\model} without the behavior-level user and item relationships, \ie, item-specific user influence ($u_i \stackrel{e\in X}{\rightarrow} v_j \stackrel{e\in X}{\rightarrow} u_{i'}$) and user-specific item dependency ($v_j \stackrel{e\in X}{\rightarrow} u_i \stackrel{e\in X}{\rightarrow} v_{j'}$).\\\vspace{-0.1in}

\item \emph{\model}-IKI: \emph{\model} without the item semantic relatedness extracted from item relation graph $G_v$, \ie, $v_j \stackrel{e\in \mathcal{R}}{\rightarrow} h \stackrel{e\in \mathcal{R}}{\rightarrow} v_{j'}$.\\\vspace{-0.1in}

\item \emph{\model}-UIKIU: \emph{\model} without the knowledge-aware dependency between users, \ie, $u_i \stackrel{e\in X}{\rightarrow} v_j \stackrel{e\in \mathcal{R}}{\rightarrow} h \stackrel{e\in \mathcal{R}}{\rightarrow} v_{j'} \stackrel{e\in X}{\rightarrow} u_{i'}$.
\end{itemize}

\begin{figure}[t]
	\centering
	%\vspace{-0.1 in}
% 	\vspace{-0.07 in}
	\subfigure[][Epinions-HR]{
		\centering
		\includegraphics[width=0.41\columnwidth]{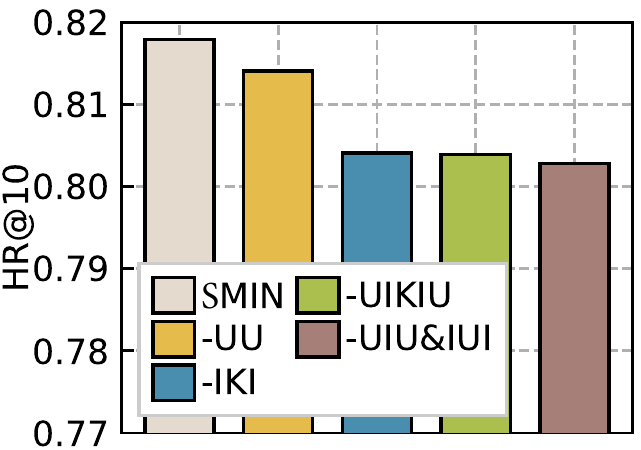}
		\label{fig:ab_yelp_NDCG}
	}
	\subfigure[][Epinions-NDCG]{
		\centering
		\includegraphics[width=0.41\columnwidth]{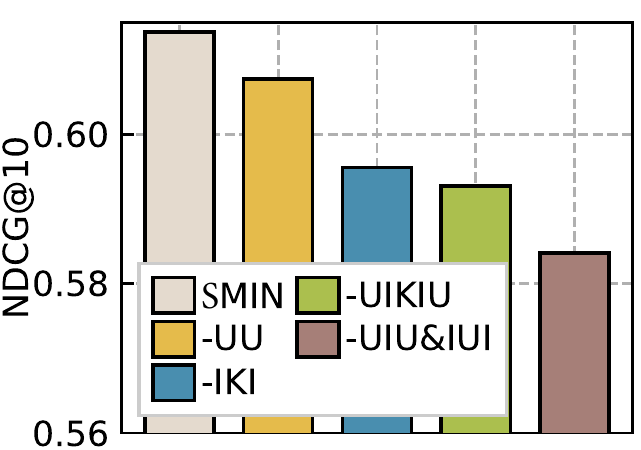}
		\label{fig:ab_yelp_NDCG}
	}
	\subfigure[][Yelp-HR]{
		\centering
		\includegraphics[width=0.41\columnwidth]{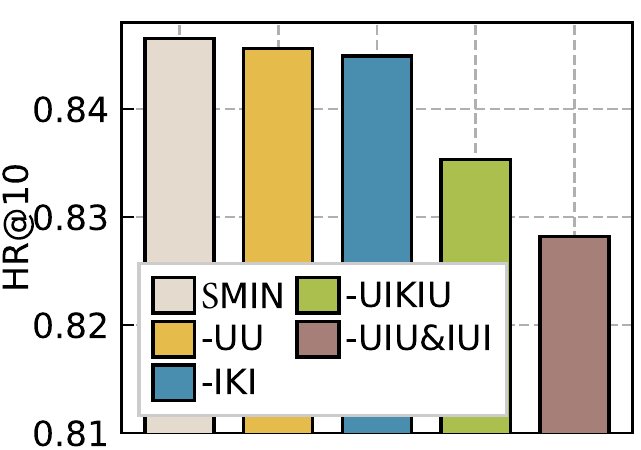}
		\label{fig:ab_ml10m_NDCG}
	}
	\subfigure[][Yelp-NDCG]{
		\centering
		\includegraphics[width=0.41\columnwidth]{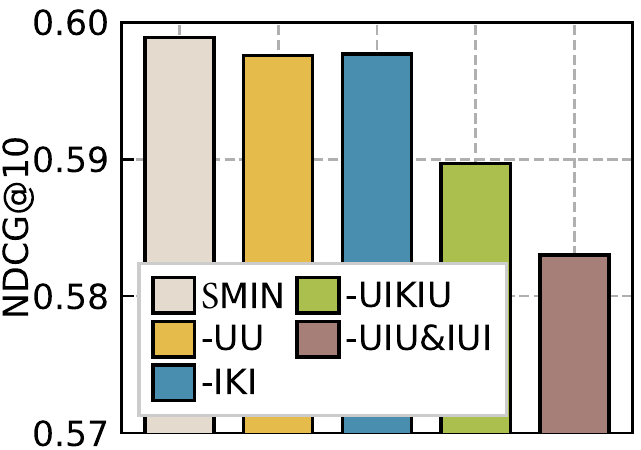}
		\label{fig:ab_ml10m_NDCG}
	}
% 	\subfigure[][E-commerce-HR]{
% 		\centering
% 		\includegraphics[width=0.29\columnwidth]{metapath_ablation_Taobao_HR@10.pdf}
% 		\label{fig:ab_retail_NDCG}
% 	}
% 	\subfigure[][E-commerce-NDCG]{
% 		\centering
% 		\includegraphics[width=0.29\columnwidth]{metapath_ablation_Taobao_NDCG@10.pdf}
% 		\label{fig:ab_retail_NDCG}
% 	}
	\vspace{-0.15in}
	\caption{Effect analyses of meta-relation ($p\in P$) incorporation in \emph{\model} in terms of \textit{HR@10}.}% and \textit{NDCG@10}.}
	\label{fig:metapath_ablation}
	%\vspace{-0.2in}
\end{figure}

The investigation results are shown in Figure~\ref{fig:metapath_ablation}, we could observe that \emph{\model} with the full set of meta-relationships obtains the best recommendation accuracy as compared to other alternatives. In particular, i) by competing with \emph{\model}-UU, social context with the network homophily has positive effects to model user's influence. ii) \emph{\model}-UIU\&IUI erases the performance gain achieved by our model. This justifies the importance of capturing item-specific user influence and user-specific item dependency, in encoding relations across users and items. iii) When competing with \emph{\model}-IKI, the results emphasizes the need of latent semantic relatedness between items in recommendation. iv) Through integrating the relational graph signals with item-specific user preference into our metapath generation phase, our scheme could further improve the recommendation accuracy as compared to \emph{\model}-UIKIU.

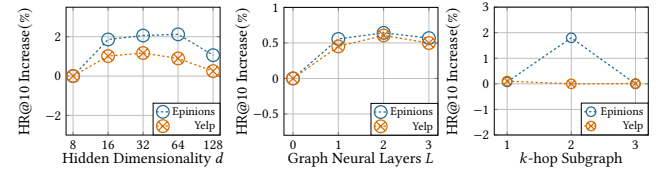
\begin{figure}
    \vspace{-0.05in}
    \centering
    \begin{adjustbox}{max width=1.0\linewidth}
    % \begin{filecontents*}{latFactor.txt}
% para  Ciao_hr   Ciao_ndcg   ep_hr   ep_ndcg    yelp_hr    yelp_ndcg ecom_hr ecom_ndcg
% 1     0.6919    0.4751      0.8014   0.5920     0.8380      0.5864  0.7489  0.5350
% 2     0.7034    0.4934      0.8163   0.6121     0.8465      0.5989  0.7546  0.5440
% 3     0.7042    0.4905      0.8179   0.6137     0.8478      0.6030  0.7551  0.5446
% 4     0.7164    0.5046      0.8184   0.6154     0.8455      0.5981  0.7588  0.5501
% 5     xxxxxx    xxxxxx      0.8099   0.6043     0.8401      0.5903  0.7497  0.5390
%row 5 set d = 128
% \end{filecontents*}
%%embedding size experiments

\begin{tikzpicture}
\begin{axis}[
    xlabel={Hidden Dimensionality $d$},
    ylabel={HR@10 Increase$(\%)$},
    xmin=0.8,xmax=5.2,
    ymin=-3.00,ymax=3.5,
    xtick={1,2,3,4,5},
    xticklabels={8,16,32,64,128},
    legend columns=1,
    legend cell align=right,
    grid=both,
    every axis plot/.append style={ultra thick},
    every tick label/.append style={scale=2},
    label style={scale=2.2},
    ylabel style={yshift=2ex},
    legend style={
        nodes={scale=1.5, transform shape},
        % legend image post style={scale=1.5},
        },
    legend style={at={(1,0)},anchor=south east},
    every axis plot post/.append style={
        every mark/.append style={scale=2}
    }
]
% \addplot[color={rgb:blue,2;green,2;yellow,1}, mark=x, dashed, mark options={solid, scale=1.5}]
% table[x=para, y=Ciao_hr] {latFactor.txt};
\addplot[color={rgb:blue,4;green,2;yellow,1}, mark=o, dashed, mark options={solid, scale=2}]
table[x=para, y=ep_hr] {latFactor.txt};
\addplot[color={rgb:red,4;green,1;yellow,2}, mark=otimes, dashed, mark options={solid, scale=2}]
table[x=para, y=yelp_hr] {latFactor.txt};
% \addplot[color={rgb:red,4}, mark=+, dashed, mark options={solid, scale=2}]
% table[x=para, y=ecom_hr] {latFactor.txt};
\legend{\large Epinions, \large Yelp};
\end{axis}
\end{tikzpicture}

\begin{tikzpicture}
\begin{axis}[
    xlabel={Graph Neural Layers $L$},
    ylabel={HR@10 Increase$(\%)$},
    xmin=0.8,xmax=4.2,
    ymin=-0.8,ymax=1,
    xtick={1,2,3,4},
    xticklabels={0,1,2,3},
    legend columns=1,
    legend cell align=right,
    grid=both,
    every axis plot/.append style={ultra thick},
    every tick label/.append style={scale=2},
    label style={scale=2.2},
    ylabel style={yshift=2ex},
    legend style={
        nodes={scale=1.5, transform shape},
        },
    legend style={at={(1,0)},anchor=south east},
    every axis plot post/.append style={
        every mark/.append style={scale=2}
    }
]
% \addplot[color={rgb:blue,2;green,2;yellow,1}, mark=x, dashed, mark options={solid, scale=1.5}]
% table[x=para, y=Ciao_hr] {latFactor2.txt};
\addplot[color={rgb:blue,4;green,2;yellow,1}, mark=o, dashed, mark options={solid, scale=2}]
table[x=para, y=ep_hr] {latFactor2.txt};
\addplot[color={rgb:red,4;green,1;yellow,2}, mark=otimes, dashed, mark options={solid, scale=2}]
table[x=para, y=yelp_hr] {latFactor2.txt};
% \addplot[color={rgb:red,4}, mark=+, dashed, mark options={solid, scale=2}]
% table[x=para, y=ecom_hr] {latFactor2.txt};
% \legend{\large Epinions, \large Yelp, \large E-commerce};
\legend{\large Epinions, \large Yelp};
\end{axis}
\end{tikzpicture}

\begin{tikzpicture}
\begin{axis}[
    xlabel={$k$-hop Subgraph},
    ylabel={HR@10 Increase$(\%)$},
    xmin=0.8,xmax=3.2,
    ymin=-2.0,ymax=3.00,
    xtick={1,2,3},
    xticklabels={1,2,3},
    legend columns=1,
    legend cell align=right,
    grid=both,
    every axis plot/.append style={ultra thick},
    every tick label/.append style={scale=2},
    label style={scale=2.2},
    ylabel style={yshift=2ex},
    legend style={
        nodes={scale=1.5, transform shape},
        },
    legend style={at={(1,0)},anchor=south east},
    every axis plot post/.append style={
        every mark/.append style={scale=2}
    }
]
% \addplot[color={rgb:blue,2;green,2;yellow,1}, mark=x, dashed, mark options={solid, scale=1.5}]
% table[x=para, y=Ciao_hr] {latFactor3.txt};
\addplot[color={rgb:blue,4;green,2;yellow,1}, mark=o, dashed, mark options={solid, scale=1.5}]
table[x=para, y=ep_hr] {latFactor3.txt};
\addplot[color={rgb:red,4;green,1;yellow,2}, mark=otimes, dashed, mark options={solid, scale=1.5}]
table[x=para, y=yelp_hr] {latFactor3.txt};
% \addplot[color={rgb:red,4}, mark=+, dashed, mark options={solid, scale=1.5}]
% table[x=para, y=ecom_hr] {latFactor3.txt};
% \legend{\arge Epinions, \large Yelp, \large E-commerce};
\legend{\large Epinions, \large Yelp};
\end{axis}
\end{tikzpicture}

% \begin{tikzpicture}
% \begin{axis}[
%     xlabel={$k$-hop Subgraph},
%     ylabel={NDCG@10 Increase$(\%)$},
%     xmin=0.8,xmax=3.2,
%     ymin=-4.50,ymax=5.5,
%     xtick={1,2,3},
%     xticklabels={1,2,3},
%     legend columns=1,
%     legend cell align=right,
%     grid=both,
%     every axis plot/.append style={ultra thick},
%     every tick label/.append style={scale=2},
%     label style={scale=2.2},
%     ylabel style={yshift=2ex},
%     legend style={
%         nodes={scale=1.5, transform shape},
%         },
%     legend style={at={(1,0)},anchor=south east},
%     every axis plot post/.append style={
%         every mark/.append style={scale=2}
%     }
% ]
% % \addplot[color={rgb:blue,2;green,2;yellow,1}, mark=x, dashed, mark options={solid, scale=1.5}]
% % table[x=para, y=Ciao_ndcg] {latFactor3.txt};
% \addplot[color={rgb:blue,4;green,2;yellow,1}, mark=o, dashed, mark options={solid, scale=1.5}]
% table[x=para, y=ep_ndcg] {latFactor3.txt};
% \addplot[color={rgb:red,4;green,1;yellow,2}, mark=otimes, dashed, mark options={solid, scale=1.5}]
% table[x=para, y=yelp_ndcg] {latFactor3.txt};
% \addplot[color={rgb:red,4}, mark=+, dashed, mark options={solid, scale=1.5}]
% table[x=para, y=ecom_ndcg] {latFactor3.txt};
% \legend{\arge Epinions, \large Yelp, \large E-commerce};
% \end{axis}
% \end{tikzpicture}
    \end{adjustbox}
    \vspace{-0.2in}
    \caption{Hyperparameter study. y-axis shows the performance increase/decrease compared to the first data point.}
    \vspace{-0.1in}
    \label{fig:hyperparam}
\end{figure}

%\vspace{-0.1in}
\subsection{Hyperparameter Study (RQ4)}
\label{sec:parameter}
To gain further insights into our \emph{\model} architecture, we now study the impact of hyperparameter settings from the following aspects. Figure~\ref{fig:hyperparam} shows the results on different datasets. We integrate evaluation results (with different value scales) of two datasets into the same figure by presenting the performance increase/decrease percentage (compared with the first data point) in y-axis.\\\vspace{-0.12in}

\noindent \textbf{Hidden Unit Dimensionality $d$}. We vary the hidden dimensionality $d$ of our latent representations from 8 to 128. The model performance saturates as the number of hidden units reaches around 64, since the larger embedding dimensionality may bring stronger representation ability. However, the further increase of $d$ leads to the slightly performance degradation due to the overfitting issue. \\\vspace{-0.12in}

\noindent \textbf{Depth of Graph Neural Encoder $L$}. We analyze the influence of the number of GNN layers by varying $L$ from 1 to 3 while keeping other parameters as default settings. The best performance of our \emph{\model} is achieved when $L=2$. Further stacking more information propagation layers (\eg, $L=3$) leads to the worse performance, which suggests that two-order connections in our heterogeneous graph architecture is sufficient to capture the multifaceted relations across users and items. \\\vspace{-0.12in}

\noindent \textbf{Substructure-aware $k$-Hops Neighbors}. We further examine the impact of incorporating $k$-hop neighboring relations into our global context injection with the awareness of graph substructure. We could notice that the performance is boosted with the increase of $k$ as 2 at the beginning stage. By modeling the global-level collaborative relations with the higher-order ($k=3$), it may introduce some noise and irrelevant dependence signals.

\begin{figure}[t]
	\centering
	%\vspace{-0.1 in}
	\subfigure[][SMIN]{
		\centering
		\includegraphics[width=0.29\columnwidth]{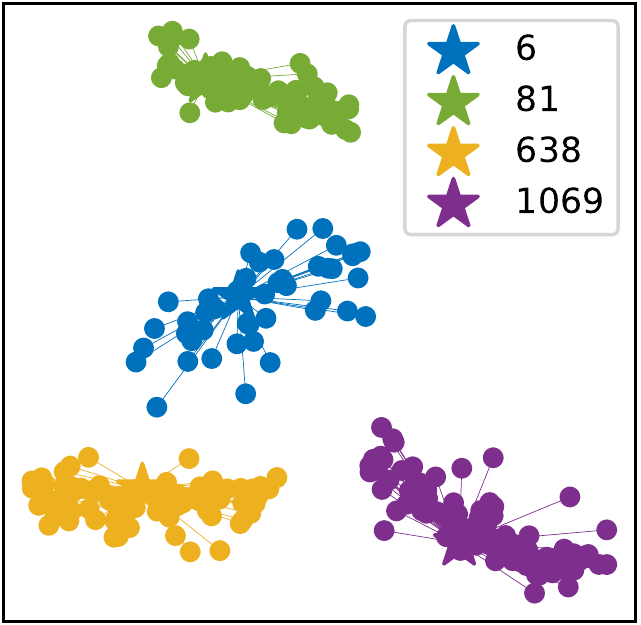}
		\label{fig:embed_NGCF}
	}
	\subfigure[][GraphRec]{
		\centering
		\includegraphics[width=0.29\columnwidth]{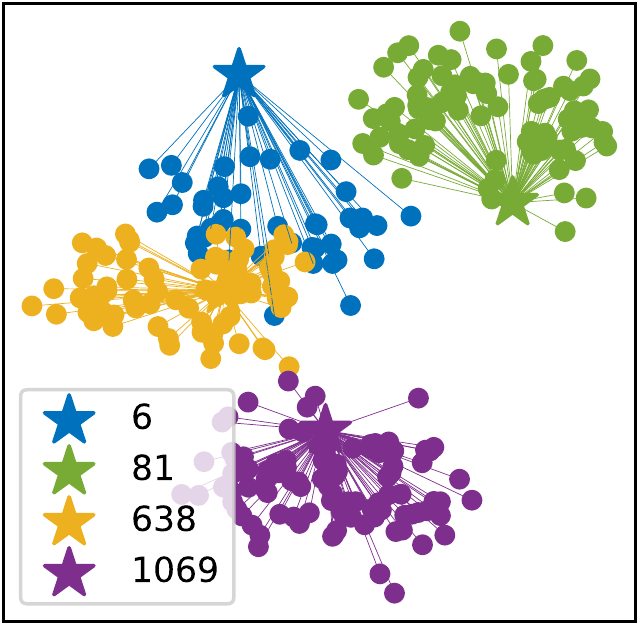}
		\label{fig:embed_KCGN}
	}
	\subfigure[][HERec]{
		\centering
		\includegraphics[width=0.29\columnwidth]{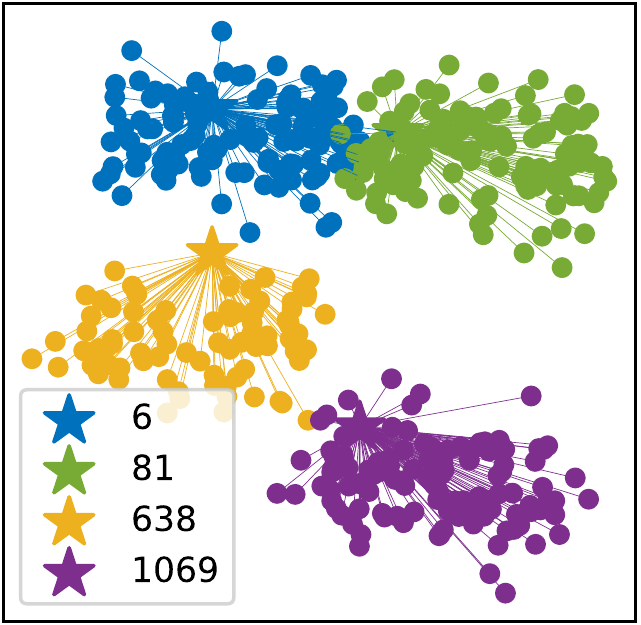}
		\label{fig:embed_KCGN}
	}
	\vspace{-0.1in}
	\caption{Embedding visualization of users (stars) and interacted items (circles) based on different encoding methods.}
	\label{fig:embed_case}
	\vspace{-0.1in}
\end{figure}

\subsection{User/Item Embedding Visualizations (RQ5)}
We perform the qualitative study on the user and item latent representations learned by \emph{\model}. In particular, we project the dense embedding vectors to 2-D space with the t-SNE algorithm~\cite{maaten2008visualizing}. As shown in Figure~\ref{fig:embed_case}, we can observe that embeddings learned by \emph{\model} present a better clustering phenomenon (denoted with the same color) compared with the embeddings of GraphRec and HERec. This suggests the effectiveness of our method in well preserving the user-item interactive relationships. Additionally, a better embedding separation phenomenon can be observed for differentiating users who interact with different items. These observations further verify the rationality of our graph-structured main embedding space which bridges the semantics between user and item domain, such that the external knowledge can be utilized to guide the heterogeneous relation representation.

\section{Related Work}
\label{sec:relate}
% In this section, we discuss the research work which is relevant to our studied problem from the following three aspects.

%\vspace{-0.05in}
\subsection{Social Recommendation Techniques}
%\noindent \textbf{Social Recommendation Techniques}.
% To alleviate sparsity of collaborative filtering models, a plenty of studies have been proposed to exploit social relationships for recommendation~\cite{fan2019graph,wu2019neural,liu2018social}. Along with this research line, earlier studies extend the matrix factorization architecture to incorporate social signals, such as SocialMF~\cite{jamali2010matrix}, SoRec~\cite{ma2008sorec} and TrustMF~\cite{yang2016social}. For example, TrustMF projects users into latent embeddings with the consideration of users' trust relationships and interaction behavior.
In recent years, many efforts have been devoted to developing social-aware recommendation techniques with various neural network architectures. In particular, attention networks has been used for social influence modeling between different users~\cite{chen2019efficient,chen2019social}. Furthermore, due to the graph-structured nature of social relations, several social recommendation methods have been proposed to handle social network information with graph neural networks~\cite{2021knowledge}. For instance, DiffNet~\cite{wu2019neural} leverages the graph convolution network to approximate the influence diffusion across users. Additionally, graph-based attentive aggregation models (\eg, DANSER~\cite{wu2019dual} and GraphRec~\cite{fan2019graph}) adapt attention mechanism to the user-user relation graph, capturing social signals for recommendation. However, most of those methods fail to incorporate the knowledge-aware information from item domain into the social-aware user-item interaction encoding. To fill this gap, this work develops a metagraph informax network, to enhance the social recommendation with the explicitly modeling of item-wise dependency information--containing fruitful facts and semantic relatedness about items.

%\vspace{-0.1in}
\subsection{GNNs for Recommendation}
%\noindent \textbf{GNNs for Recommendation}.
The objective of graph neural networks is to project nodes in a generated graph into a low-dimensional vector space~\cite{wang2019heterogeneous,zhang2019heterogeneous}. The rationality which motivates the development of graph neural network, is that nodes are naturally characterized by their own features and neighbors~\cite{qiu2018network,xia2021graph,huangrecent,fan2019metapath}. Following this idea, many recent efforts aim to explore the user-item interaction graph structure for recommendation scenarios~\cite{xia2021knowledge}. For example, based on the graph convolution operation, several studies propose to capture the collaborative relationships between users and items over their interaction graph, such as PinSage~\cite{ying2018graph} and NGCF~\cite{wang2019neural}. Later on, LightGCN~\cite{he2020lightgcn} optimizes the GCN-based user-item relation encoder by only keeping the most essential module--neighborhood aggregation in the graph-structured message passing paradigm. This work develops a hierarchical learning framework by integrating a metapath-guided graph neural module and a substructure-aware mutual information modeling paradigm for social-aware recommender system.

% For example, DKN~\cite{wang2018dkn} introduces a convolution-based model to make recommendations on news with the combination of entity and text embeddings. A knowledge-guided sequential recommendation method is developed to consider sequence- and knowledge-level rewards based on the markov decision process~\cite{wang2020kerl}.

%\vspace{-0.1in}
% \subsection{Knowledge-aware Recommender Systems}
% Another relevant research line of recommendation methods lies in incorporating the knowledge graph information as the side information of items~\cite{zhu2020knowledge}. These knowledge-aware recommender systems transfer structural knowledge among items to the user-item interaction modeling to improve the recommendation performance~\cite{xian2019reinforcement,chen2020jointly}. With the promising results of graph neural networks in relation learning, researchers have recently attempted to investigate the structural knowledge with the embedding propagation paradigm, such as KGAT~\cite{wang2019kgat} and KGNN-LS~\cite{wang2019knowledge}. Different from these knowledge-aware recommendation methods, this paper empower recommender systems with the ability of utilizing not only the knowledge information from item domain, but also exploring the multifaceted social signals, in order to boost the performance.

\subsection{Self-Supervised Learning}
%\noindent \textbf{Self-Supervised Learning}.
Self-supervised learning has shown its effectiveness in learning representations from limited labeled data, such as natural language processing~\cite{rubino2020intermediate,huang2021ghostbert,ruiter2020self} and image data analysis~\cite{misra2020self,mustikovela2020self}. The general idea of self-supervised learning is to design pretext training tasks, so as to offer additional supervision signals. There exists one line of recent trend aims to propose self-supervised learning paradigms over graph-structured data~\cite{yu2021self,2021graphsession,hu2020gpt}. For example, Hu~\etal~\cite{hu2020gpt} proposes a pre-trained graph neural network based on the graph reconstruction task in terms of the attributed and structural information. Additionally, a hypergraph convolutional network is augmented with a self-supervised learning framework with respect to the mutual information exploration~\cite{yu2021self}. Different from the above methods, this work is the first to inject the relational knowledge from user and item side as the self-supervised information into social recommendation. This will enhance the user/item representation paradigm to learn a knowledge-aware global information, which leads to better recommendation performance.

% For example, DKN~\cite{wang2018dkn} introduces a convolution-based model to make recommendations on news with the combination of entity and text embeddings. A knowledge-guided sequential recommendation method is developed to consider sequence- and knowledge-level rewards based on the markov decision process~\cite{wang2020kerl}.

%\vspace{-0.1in}
\subsection{Knowledge-aware Recommender Systems}
Another relevant research line of recommendation methods lies in incorporating the external knowledge information as the side information of items~\cite{zhu2020knowledge,xia2021knowledge}. These knowledge-aware recommender systems transfer structural knowledge among items to the user-item interaction modeling to improve the recommendation performance~\cite{xian2019reinforcement,chen2020jointly}. With the promising results of graph neural networks in relation learning, researchers have recently attempted to investigate the structural knowledge with the embedding propagation paradigm, such as KGAT~\cite{wang2019kgat} and KGNN-LS~\cite{wang2019knowledge}. Different from these knowledge-aware recommendation methods, this paper empower recommender systems with the ability of utilizing not only the knowledge information from item domain, but also exploring the multifaceted social signals, in order to boost the performance.

\section{Conclusion}
\label{sec:conclusion}

In this paper, we propose the Self-Supervised Metagraph Infomax Networks (\model) architecture for knowledge-aware social recommendation. Our model learns social and knowledge graph dependent relationships for users and items via a metapath-guided heterogeneous graph neural network. Furthermore, a self-supervised learning framework is introduced to augment the modeling of graph structural information, so as to preserve the complex relationships among users and items. Extensive experimental results on several real-life datasets demonstrate the advantage of our framework as compared to various state-of-the-art recommendation models.

% Our future work involves exploring user and item external attributes (\eg, user profile, product images) to further improve the performance.

%Furthermore, we plan to deploy \model\ in a real-time recommendation scenario.% It is important to extend our knowledge-enhanced social recommender systems to be well suited to handling steaming data.

% we plan to endow \model\ with the ability of handling data stream of user behavior with the parameter optimization in a real-time manner.

\section*{Acknowledgments}
We thank the reviewers for their valuable comments. This work is supported by National Nature Science Foundation of China (62072188), Major Project of National Social Science Foundation of China (18ZDA062), Science and Technology Program of Guangdong Province (2019A050510010).

\clearpage

\bibliographystyle{ACM-Reference-Format}
\bibliography{sigproc}

\end{document}